\documentclass{aastex}
\usepackage{emulateapj5}
\usepackage{apjfonts}
\usepackage{epsf}

\usepackage{graphics}

\newcommand{\be}{\begin{equation}}
\newcommand{\ee}{\end{equation}}

\slugcomment{accepted by ApJ Letters}

\shorttitle{Evidence for the evolutionary sequence of blazars}
\shortauthors{Cao X}

\begin{document}

\title{Evidence for the evolutionary sequence of blazars:
different types of accretion flows in BL Lac objects}

\author{Xinwu Cao}
\affil{Shanghai Astronomical Observatory, Chinese Academy of Sciences,
Shanghai, 200030, China\\
Email: cxw@center.shao.ac.cn}


\begin{abstract}
The limits on the mass of the black hole in 23 BL Lac objects
are obtained from their luminosities of the broad emission line
H$\beta$ on the assumption that broad emission lines are emitted from
clouds ionized by the radiation of the accretion disk surrounding a black
hole. 
The distribution of line luminosity $L_{{\rm H}\beta}$ of all these BL Lac objects 
suggests a bimodal nature, although this cannot be statistically proven on the
basis of the present, rather small sample. 
We found that standard thin disks are
probably in the sources with $L_{{\rm H}\beta}>10^{41}$ erg~s$^{-1}$.
The central black holes in these sources have masses of
$10^{8-10} M_\odot$, if the matter is accreting at the rate of
$0.025 {\dot M}_{\rm Edd}$. For the sources with
$L_{{\rm H}\beta}<10^{41}$ erg~s$^{-1}$, the accretion flows have transited 
from standard thin disk type to the ADAF type.  The lower limits on the
mass of the black hole in these sources are in the range of
$1.66-24.5\times 10^{8}$ $ M_\odot$. The results support the
evolutionary sequence of blazars: FSRQ$\rightarrow$LBL$\rightarrow$HBL.
\end{abstract}


\section{Introduction}

The masses of the central black holes in quasars are crucial in
understanding the evolution of quasars. Some different approaches are
proposed to estimate the masses of the black holes in active galactic
nuclei (AGNs), such as the gas kinematics near a hole (see \citet{hk00}
for a review and references therein). The black hole mass
derived from the direct measurements on the gases moving near the hole
is reliable, but it is only available for very few AGNs. For most AGNs,
the central black hole masses can be inferred from the velocities of the
clouds in broad line regions (BLRs) and the sizes of BLRs
\citep{d81,wpm99,kas00}. 
Most BL Lac objects have featureless optical and ultraviolet continuum
spectra, and only a small fraction of BL Lac objects show very weak
broad emission lines \citep{vv00}. It is therefore
difficult to estimate their central black hole mass from the kinematics
of their BLRs. The discovery of a tight correlation between stellar velocity
dispersion and black hole mass in nearby galaxy bulges
offers a new tool for estimate of
the black hole mass \citep{fm00, g00}. The black hole mass can be estimated from the
measurement of bulge velocity dispersion using the $M-\sigma$
relation. The masses of the black holes in some nearby BL Lac objects
have been estimated from the measurement of their bulge velocity
dispersion
\citep{fkt01,fkt02,bhs02}.
\citet{wu02} adopted the fundamental plane for ellipticals to estimate
the velocity dispersion and then the hole masses for some AGNs. 
\citet{fan} derived the masses of the black holes
in Mkn 501 and some other blazars from the gamma-ray variability time scale.
They found that the masses are around $10^7 M_{\odot}$.

\citet{gc98} used a large sample of blazar broadband
spectra to study the blazar sequence. They suggested a sequence:
HBL$\rightarrow$LBL$\rightarrow$FSRQ. This sequence represents an
increasing energy density of the external radiation field that
leads to an increasing amount of Compton cooling. The decrease
of the maximum energy in the electron distribution causes the synchrotron
and Compton peaks to shift to lower frequencies.
\citet{gkm01} argued that the radiating jet plasma is outside the
broad line scattering region in weak sources and within it in
powerful sources, and the model fits to the spectra of several blazars
proposed a sequence: FSRQ$\rightarrow$LBL$\rightarrow$HBL. 
The evolutionary  sequence:  FSRQ$\rightarrow$LBL$\rightarrow$HBL, has
recently been suggested by \citet{dc00}  and \citet{cd01}. In this
evolutionary sequence, less gas is left to
fuel the central engine for BL Lac objects, and the advection dominated
accretion flows (ADAFs) may be in most BL Lac objects. The blazar
spectral calculations support this scenario \citep{bd02}.

There are several tens of BL Lac objects in which one (or more) broad
emission line has been detected. It is therefore possible to infer
the central ionizing luminosity
through their broad line emission for these BL Lac objects.
The limits on the central black hole mass can be obtained, if the
accretion type in the central engine is known. The cosmological
parameters $H_{0}=75$ kms$^{-1}$ Mpc$^{-1}$ and $q_{0}=$0.5 have
been adopted in this work.

\section{Estimate on the ionizing luminosity}

It is not possible to measure the ionizing luminosity directly from
observations on BL Lac objects, since the observed continuum emission
from the jets is strongly beamed to us. In this case, the optical
emission line luminosity can be used to estimate the central
ionizing luminosity \citep{rs91,c97,cj99}.

We can estimate the ionizing continuum luminosity $L_{\lambda, \rm ion}$
at the given wavelength $\lambda_{0}$ as

\be
L_{\lambda, \rm ion}(\lambda_0)={\frac {L_{\rm line}}{EW_{\rm ion}}},
\ee
where $\lambda_0$ is the wavelength of the line,
$EW_{\rm ion}$ is the equivalent width of the broad emission line
corresponding to the ionizing continuum emission (different from the
observed continuum emission).

The uncertainty in Eq. (1)  is the value of $EW_{\rm ion}$, which may be 
different for individual sources. We estimate the value
of $EW_{\rm ion}$ from the \citet{bg92} sample.
This sample contains all 87 PG quasars ($z<0.5$) with high quality
optical spectra. The average value of $EW_{{\rm H}\beta}$ is
100 \AA ~ for 70 radio-quiet quasars. Unlike blazars, the optical continuum of
radio-quiet quasar has not been contaminated by the beamed synchrotron
emission from the jet. We will take $EW_{\rm ion}=100$ \AA ~  and use 
the broad emission line ${\rm H}\beta$ to estimate the ionizing continuum
luminosity of BL Lac objects. The BL Lac objects seem to follow the
statistical behavior of quasars in the correlations between radio
and broad line emission \citep{c97,cj99}. It may imply that the properties
of BLRs in BL Lac objects are not significantly different from that
in quasars. If the $EW_{\rm ion}$ of BL Lac objects deviates from that of
quasars systematically, then the estimated black hole mass could be
modified with $EW_{\rm ion}$ (see further discussion in Sect. 5).

\section{ Estimate of the black hole mass}

For a standard thin disk, the ionizing continuum luminosity is
mainly determined by the central black hole mass and accretion rate.
For a low accretion rate, the accretion flow will transit from standard
thin disk type to the ADAF type \citep{ny95,y96}.

\subsection{Standard thin accretion disks}

The standard thin accretion disks are thought to be in most quasars
\citep{kb99}. In this case, the luminosity at optical
wavelength can be related with the disk luminosity by 
$L_{\rm d}\simeq  9\lambda L_{\lambda, {\rm ion}}$(5100  \AA)
\citep{kas00}. The
central black hole mass $M_{\rm bh}$ is then estimated by
$M_{\rm bh}\simeq 10^{-38} (L_{\rm d}/{\dot m}) M_\odot$, 
where $\dot m=\dot M/\dot M_{\rm Edd}$. The ionizing luminosity
$L_{\lambda,\rm ion}$ can be inferred from $L_{\rm H\beta}$, and we can
finally estimate the black hole mass from the broad line luminosity
$L_{{\rm H}\beta}$, if $\dot m$ is known.
For $\dot m=1$, the lower limit of the black hole mass is available.

\subsection{ADAFs}

The transition of the accretion flow from the thin disk type to the
ADAF type occurs while $\dot m$ decreases to a value below ${\dot
m}_{\rm crit}$ \citep{ny95,y96}. One possible mechanism for 
the transition might be the evaporation of the disk
\citep{mm94,l99,rc00,mm01}.
\citet{gl00} suggested that the transition can be triggered by the
thermal instability of a radiation pressure-dominated standard
accretion disk.

The spectrum of an ADAF: $L_{\lambda}(M_{\rm bh}, {\dot m}, \alpha, \beta)$
can be calculated if the parameters $M_{\rm bh}$, $\dot m$, $\alpha$,
and the fraction of the magnetic pressure $\beta$ are specified.
The parameter $\beta$ is defined as $p_{\rm m}=(1-\beta)p_{\rm tot}$.
We can calculate the spectra of ADAFs using the approach 
proposed by \citet{m97}.  For the fixed black hole mass
$M_{\rm bh}$, the optical luminosity $L_{\lambda}$ increases
with $\alpha$. For AGNs, the viscosity $\alpha$ could be as high as 1,
as suggested by \citet{n96}. In this work, We set $\alpha=0.3$
to calculate optical luminosity $L_{\lambda}$ \citep{y96,c01}.
The fraction of magnetic pressure $\beta=0.5$ for
equipartition cases. In fact, our numerical results show that the maximal
optical luminosity $L_{\lambda}^{\rm max}$ always requires $\beta=0.5$,
if all other parameters are fixed. The accretion rate
$\dot m<\dot m_{\rm crit} \simeq 0.28\alpha^2$ should be satisfied
for an ADAF \citep{m97}. So, varying the accretion rate $\dot m$,
we can find the maximal optical luminosity at $\lambda_0$ numerically
for $\alpha=0.3$ and $\beta=0.5$ (see Fig. 1). We plot the maximal
value of $\lambda L_{\lambda}^{\rm max}(\lambda_0)$ varying with 
black hole mass $M_{\rm bh}$ in Fig. 2. We can obtain a lower limit
on the mass of the black hole from $L_{{\rm H}\beta}$ using the relation
$\lambda L_{\lambda}^{\rm max}-M_{\rm bh}$ plotted in Fig. 2.

For BL Lac objects, most accretion power may be carried by strong jets
and the accretion flows may probably be described by the adiabatic inflow-outflow
solutions (ADIOSs) \citep{bb99}. In this case, the flow is fainter than
the pure ADAF considered here. The maximal
$\lambda L_{\lambda}^{\rm max}(\lambda_0)$ is therefore still valid for
an ADIOS. It will not affect our estimate of the black hole mass limits.

\section{Masses of black holes in BL Lac objects}

\citet{d01} complied a sample of blazars observed in the X-ray band,
which includes almost all HBLs, IBLs, and many LBLs. We collect all
BL Lac objects in their sample and 1~Jy+S4+S5 catalogues
\citep{s94,sk94,sk96} as our start sample.
This sample includes most identified LBLs and HBLs.
We search the literature for all sources with broad emission line
fluxes, and this leads to a sample of 23 sources (listed in Table 1).
Most of them are LBLs, except three HBLs: 0651$+$428, Mkn 421,
and Mkn 501. The sources with relatively strong broad emission lines
($EW>10$ \AA) are considered as genuine HPQs by \citet{vv00} and
are therefore not included in our sample. The source 3C 279 (1253$-$055)
is classified as a BL Lac object in \citet{vv00} due to
its small broad line equivalent width, though this source is usually
regarded as a quasar in other literature. 

Many sources in this sample have detected Mg\,{\sc ii} or H$\alpha$
line emission instead of H$\beta$. We use line ratios reported by
\citet{f91} to estimate H$\beta$ line luminosity
$L_{{\rm H}\beta}$ from the line luminosity of H$\alpha$ or Mg\,{\sc ii}.
The broad emission line luminosity $L_{{\rm H}\beta}$ are listed
in Table 1. The distribution of the line luminosity $L_{{\rm H}\beta}$
is plotted in Fig. 3. 

We use Eq. (1) and the line luminosity $L_{{\rm H}\beta}$ to estimate
the ionizing continuum luminosity at 4861 \AA. The central
black hole masses $M_{\rm bh,1}$ and $M_{\rm bh,2}$ can be estimated
in the cases of a thin disk and an ADAF, respectively. The mass
derived for the standard thin disk case also depends on accretion rate
$\dot m$.  The lower limit of $M_{\rm bh,1}$ can be available
by setting $\dot m=1$. If the transition of the accretion flow from the
thin disk type to the ADAF type occurs at
$\dot m\sim \dot m_{\rm crit}$, we can have an upper limit on
$M_{\rm bh,1}$ setting $\dot m=0.025$ for $\alpha=0.3$.
For ADAFs, $M_{\rm bh,2}$ is the lower limit, since the maximal optical
continuum luminosity is calculated for the given black hole mass.
We list the derived black hole masses in Columns (6) and (7) of
Table 1.

\figurenum{1}
\centerline{\includegraphics[angle=0,width=8.5cm]{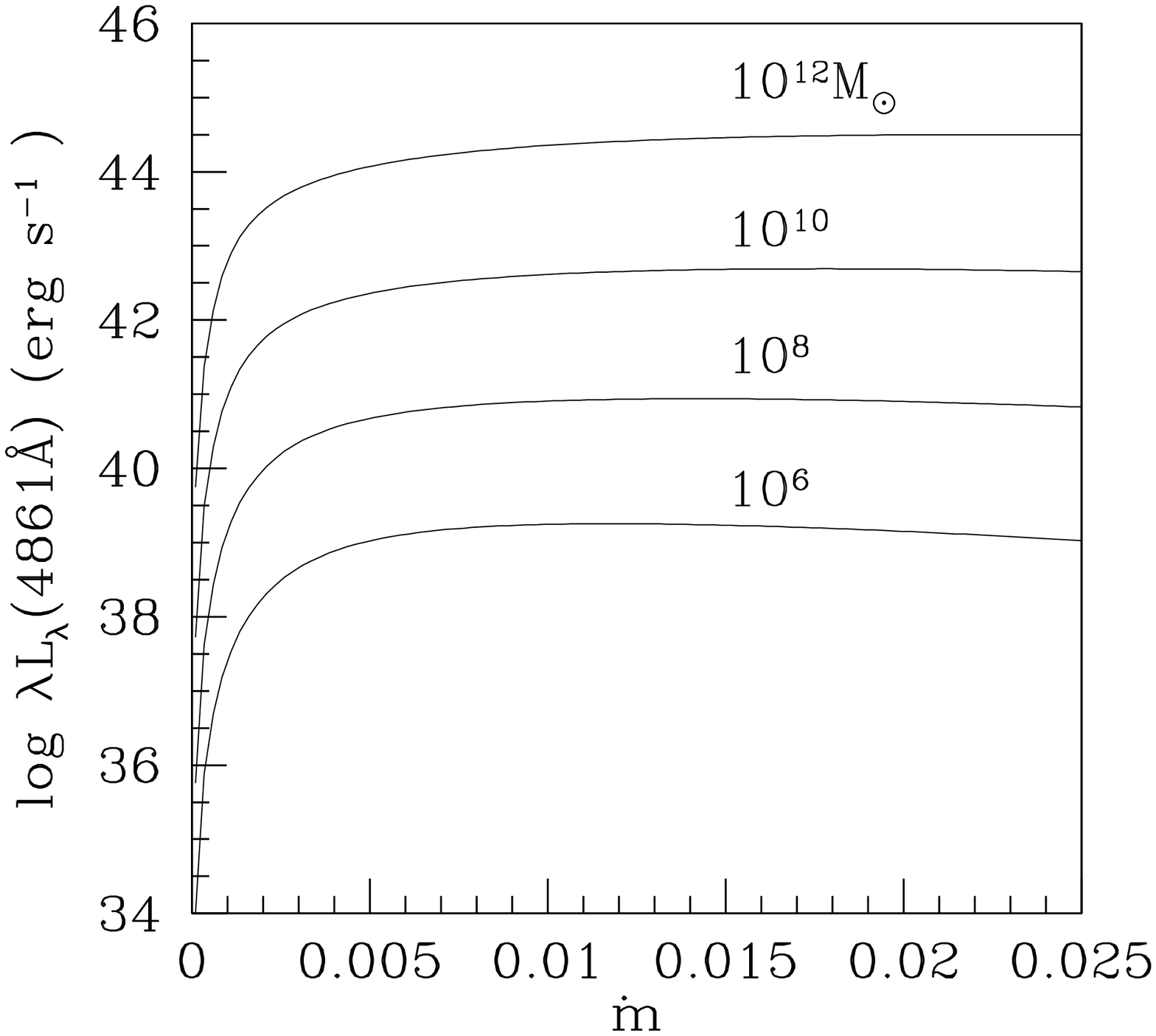}}
\figcaption{\footnotesize
The optical luminosity of ADAFs at 4861 \AA ~ varies
with accretion rate $\dot m$ for different black hole masses
($\alpha=0.3$ and $\beta=0.5$ are adopted).
\label{fig1}}
\centerline{}

\figurenum{2}
\centerline{\includegraphics[angle=0,width=8.5cm]{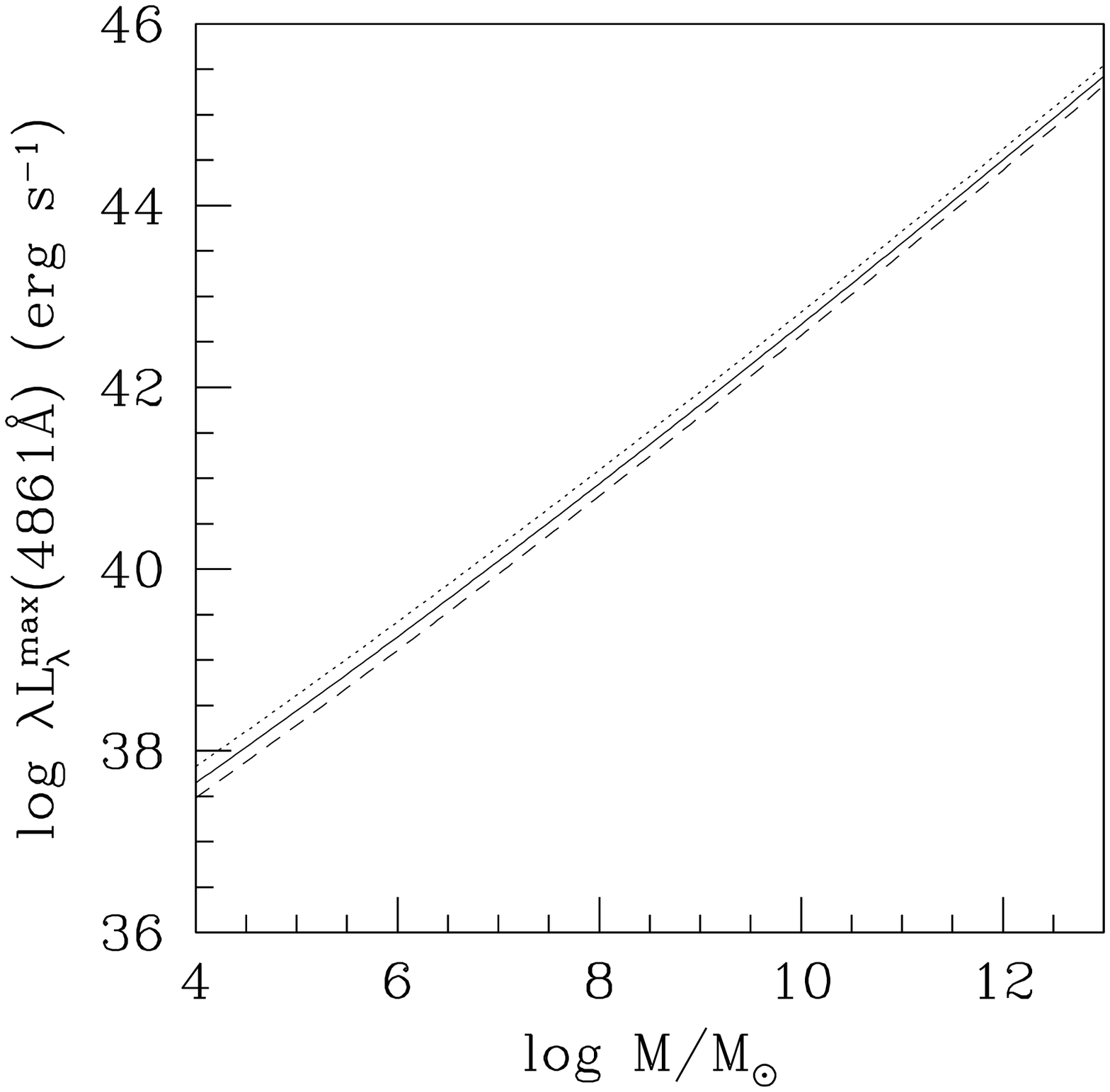}}
\figcaption{\footnotesize
The maximal optical luminosity of ADAFs at 4861 \AA ~
as functions of black hole mass for different values $\alpha$:
$\alpha=$0.1 (dashed line), 0.3 (solid), and 1 (dotted).
\label{fig2}}
\centerline{}

\figurenum{3}
\centerline{\includegraphics[angle=0,width=8.5cm]{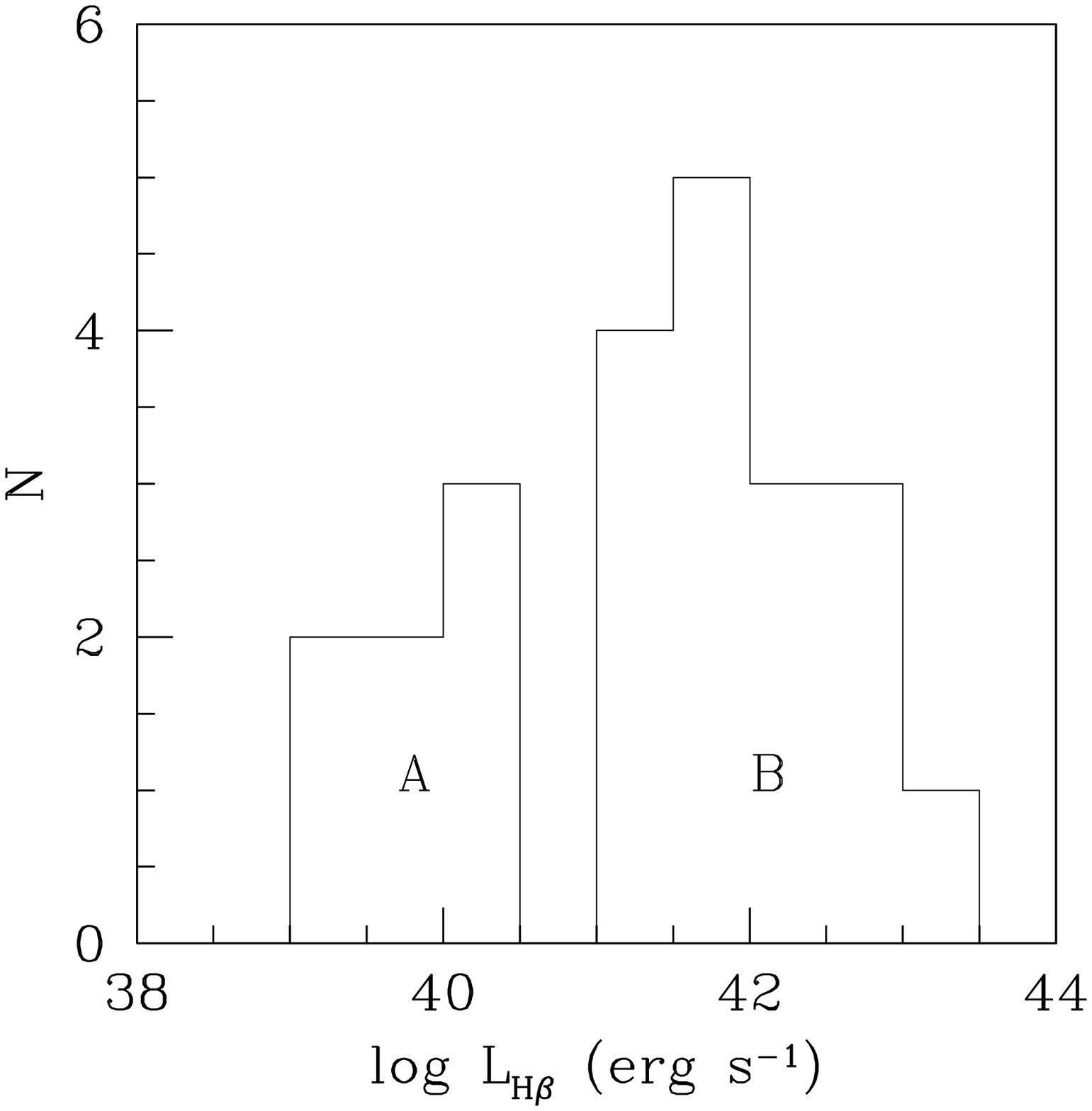}}
\figcaption{\footnotesize
The distribution of the broad emission line luminosity
$L_{\rm H{\beta}}$. 
\label{fig3}}
\centerline{}

\section{Discussion}

The distribution of line luminosity $L_{{\rm H}\beta}$ of all these
BL Lac objects suggests a bimodal nature, although this cannot be
statistically proven on the basis of the present, rather small sample
(see Fig. 3).
We define the sources with $L_{{\rm H}\beta}<10^{41}$ ergs~s$^{-1}$ as population A,
and all others are in population B. 
It is still not clear whether this bimodal nature of line luminosities 
would hold for a larger sample. It would be interesting
to perform high-sensitivity optical observations on a large sample of
BL Lac objects. 

For thin disk cases, the lower limits on the hole mass would be  
in  $10^{4-8}$~$M_\odot$, while the upper limits on the hole mass
would be in $10^{6-10}$~$M_\odot$ if $\dot m\sim 0.025$. For accretion rate $\dot m <0.025$, the accretion
flow would be in ADAF state. In this case, the lower limits on the black
hole mass are in:  $10^{8-12}$~$M_\odot$, which is similar to the
results of \citet{wang01}. Noting that this is the lower
limit, the black hole mass could be much higher if the accretion rate
is low or/and the viscosity $\alpha$ is small.

The masses of black holes in all sources of population B are in
$10^{6-8}$~$M_\odot$ if $\dot m\sim 1$. It would be interesting to
pay much attention on the seven sources in population A. 
If the accretion flows in these sources are in thin disk state,
their central black holes would have masses $10^{4-6}$~$M_\odot$ for
any value of $\dot m$. If the accretion rate $\dot m$ is 
lower than the critical value $\dot m_{\rm crit}$, the accretion flows
in these sources are in the ADAF state. The lower limits on the mass of
the black hole in these sources are in the range of
$1.66-24.5\times 10^{8}$ $ M_\odot$. \citet{fkt01} found that the mass
of the black hole in Mkn 501 is $3.2\times 10^8$~$M_\odot$ from the
measurement of bulge velocity dispersion, while \citet{bhs02}
derived the mass of $(0.9-3.4)\times 10^9$~$M_\odot$.
Considering that our estimate ($1.66\times 10^{8}$ $ M_\odot$) is the
lower limit, the central black hole mass can be higher than the limit
if the accretion rate is sufficiently low or/and the viscosity
$\alpha$ is small. Our result is consistent with their results of
Mkn 501 derived from bulge velocity dispersion. It may probably that
the sources in population A have already been in ADAF state.

The sources in population B may  have standard thin accretion disks
surrounding the black holes, otherwise some black holes should
be at least as huge as $10^{12} M_\odot$. If the accretion rate $\dot m$ of these
sources is as small as $\sim 0.025$, slightly higher than the critical
value below which the accretion flow would be in ADAF state,
the black hole mass would be in $10^{8-10}$~$M_\odot$.
As the fact that most FSRQs have black hole mass higher than
$10^8$~$M_\odot$ \citep{laor00,md01,g01},
this is compatible with the unified models of
radio-loud quasars \citep{up95}.

There are three HBLs in our sample in population A, and all sources
in population B are LBLs. It may imply that the accretion flows in all
HBLs are in ADAF state. The fact that no broad emission line has been 
detected for most BL Lac objects may imply that only a small fraction
of LBLs have optically thick standard thin accretion disks surrounding
the black holes with very low accretion rate close to
$\dot m_{\rm crit}$. We speculate that these LBLs
will finally exhaust the gas near the hole and the disks will transit 
to ADAFs. Most other LBLs and HBLs without any broad emission line
detected may be in population A, and the accretion flows have already
been in ADAF state. Otherwise the black holes in these sources should be very small,
if the accretion flows are in standard thin disk state. 
It is most probably that the BL Lac objects studied here with one
(or more) broad emission line detected are in the intermediate state
of the evolutionary sequence from FSRQ to BL Lac object.
The fact that no HBL is in population B may imply that the
evolutionary sequence of BL Lac objects should be LBL$\rightarrow$HBL.
The results present in this {\it Letter} support the evolutionary
sequence FSRQ$\rightarrow$LBL$\rightarrow$HBL suggested by
\citet{dc00}. If this is the case, then the ratio of
the BL Lac objects in population B to the remainder offers a clue to
study the detailed evolutionary history of blazars, and it can also
be a useful test on ADAF models.

If the evolution of blazars is really regulated by the gas near the
black hole, the reprocessing optical depth of the BLR would decrease
with the depletion of the gas near the hole, and the line $EW_{\rm ion}$
of the blazar would also
decrease along the evolutionary sequence. The lower limits on the mass
of the hole in evolving blazars should be modified
with varying $EW_{\rm ion}$. In this work, we used a single $EW_{\rm ion}$  to
derive the masses of holes in blazars and found that the holes have
similar masses for these two populations of BL Lac objects. 
In this case, the lower limits on the mass of the hole in
the blazars evolving at a later stage would become systematically
higher than that derived
in this work. The derived hole masses seem to be consistent with the
evolutionary sequence FSRQ$\rightarrow$LBL$\rightarrow$HBL.

\acknowledgments

I thank the anonymous referee's helpful comments. This work is support by
NSFC(No. 10173016) and the NKBRSF (No. G1999075403). This research has
made use of the NASA/IPAC Extragalactic Database (NED), which is operated by the Jet
Propulsion Laboratory, California Institute of Technology,
under contract with the National Aeronautic and Space Administration.


\begin{deluxetable}{ccccccc}
\tabletypesize{\scriptsize}
\tablecaption{Broad emission line data and black hole masses
}
\tablewidth{0pt}
\tablehead{
\colhead{Source} &
\colhead{Redshift}   &
\colhead{Line}   &
\colhead{log $L_{{\rm H}\beta}$} &
\colhead{References} &
\colhead{log {M$_{\rm bh,1}$}$^{\rm a}$ }  &
\colhead{log {M$_{\rm bh,2}$}$^{\rm b}$ }
}
\startdata
 0235$+$164   &  0.940  & Mg\,{\sc ii} & 42.2  & CJ99 &  8.30   & 11.30 \\
 0537$-$441   &  0.896  & Mg\,{\sc ii} & 43.4  & CJ99 &  9.58   & 12.60 \\
 0651$+$428$^{\rm c}$   &  0.126  & H$\alpha$  & 40.5 & M96 &  6.59 & 9.39  \\
 0814$+$425   &  0.258  & Mg\,{\sc ii} & 40.3  & CJ99 &  6.38   &  9.15 \\
 0820$+$225   &  0.951  & Mg\,{\sc ii} & 41.4  & CJ99 &  7.51   & 10.43 \\
 0823$+$033   &  0.506  & Mg\,{\sc ii} & 41.8  & CJ99 &  7.90   & 10.86 \\
 0851$+$202   &  0.306  & H$\beta$ & 41.3   & CJ99 &  7.44   & 10.35 \\
 0954$+$658   &  0.367  & H$\alpha$ & 41.0  & CJ99 &  7.16   & 10.03 \\
 1101$+$384$^{\rm c}$   &  0.031  & H$\alpha$ & 39.9 & C00  &  6.05   &  8.78\\
 1144$-$379   &  1.048  & Mg\,{\sc ii} & 42.7 & CJ99  &  8.80   & 11.85 \\
 1253$-$055   &  0.536  & H$\beta$ & 42.9 & CJ99  &  9.07   & 12.14 \\
 1400$+$162   &  0.244  & H$\beta$ & 40.5 & SJ85  &  6.59   &  9.39 \\
 1538$+$149   &  0.605  & Mg\,{\sc ii} & 41.5 & CJ99  &  7.60   & 10.52 \\
 1652$+$398$^{\rm c}$   &  0.0337 & H$\alpha$ & 39.4 & C00  &  5.57   &  8.22 \\
 1722$+$401   &  1.049  & Mg\,{\sc ii} & 41.5 & VV00  &  7.63   & 10.55 \\
 1749$+$096   &  0.320  & Mg\,{\sc ii} & 42.2 & VV00  &  8.28   & 11.28 \\
 1803$+$784   &  0.684  & H$\beta$ & 42.9 & CJ99  &  9.99   & 12.06\\
 1807$+$698   &  0.051  & H$\beta$ & 39.9 & SJ85  &  6.03   &  8.74 \\
 1823$+$568   &  0.664  & H$\beta$ & 41.7 & CJ99  &  7.80   & 10.74 \\
 1921$-$293   &  0.352  & H$\beta$ & 41.9 & JB91 &  8.04   & 11.01\\
 2029$+$121   &  1.215  & Mg\,{\sc ii} & 42.2 & CJ99  &  8.29   & 11.29 \\
 2200$+$420   &  0.068  & H$\alpha$ & 39.5 & SJ85  &  5.63   &  8.28 \\
 2240$-$260   &  0.774  & Mg\,{\sc ii} & 41.8 & CJ99  &  7.98   & 10.94 \\
 \enddata
\tablenotetext{a}{in unit of 0.025M$_{\odot}/{\dot m}$}
\tablenotetext{b}{in unit of M$_{\odot}$}
\tablenotetext{c}{HBL}
\tablerefs{C00: \citet{cao00}; CJ99: \citet{cj99};
JB91: \citet{jb91}; M96: \citet{m96}; SJ85: \citet{sj85}; VV00:
\citet{vv00}.
}
\end{deluxetable}




\begin{thebibliography}{}
\bibitem[Barth, Ho, \& Sargent(2002)]{bhs02}Barth, A.J., Ho, L.C., \&
   Sargent, W.L.W., 2002, \apj, in press (astro-ph. 0201064)
\bibitem[Blandford, \& Begelman(1999)]{bb99}
        Blandford, R.D., \& Begelman, M.C., 1999, \mnras, 303, L1
\bibitem[Boroson \& Green(1992)]{bg92}
       Boroson, T.A., \& Green, R.F., 1992, \apjs, 80, 109
\bibitem[B\"ottcher \& Dermer(2002)]{bd02}
       B\"ottcher, M., \& Dermer, C.D., 2002, \apj, 564, 86
\bibitem[Cao(2000)]{cao00}
       Cao, X., 2000, \aap, 355, 44
\bibitem[Cao \& Jiang(1999)]{cj99}
       Cao, X., \& Jiang, D.R., 1999, \mnras, 307, 802
\bibitem[Cavaliere \& D'Elia(2001)]{cd01}Cavaliere, A.,
        \& D'Elia, V., 2001, \apj, submitted (astro-ph. 0106512)
\bibitem[Celotti, Padovani, \& Ghisellini(1997)]{c97}
       Celotti, A., Padovani, P., \& Ghisellini, G., 1997, \mnras, 286, 415
\bibitem[Choi, Yang, \& Yi(2001)]{c01}
     Choi, Y.-Y., Yang, J., \& Yi, I., 2001, \apj, 555, 673
\bibitem[D'Elia \& Cavaliere(2000)]{dc00}
        D'Elia, V., \& Cavaliere, A., 2000, \pasp, 227, 252
\bibitem[Dibai(1981)]{d81}Dibai, E.A., 1981, Soviet Astron., 24, 389
\bibitem[Donato et al. (2001)]{d01}
     Donato, D., Ghisellini, G., Tagliaferri, G., \& Fossati, G.,
     2001, \aap, 375, 739
\bibitem[Falomo, Kotilainen, \& Treves (2001)]{fkt01}Falomo, R.,
     Kotilainen, J., \& Treves, A., 2001, astro-ph. 0112423
\bibitem[Falomo, Kotilainen, \& Treves (2002)]{fkt02}Falomo, R.,
     Kotilainen, J., \& Treves, A., 2002, \apj, in press
     (astro-ph. 0203199)
\bibitem[Fan, Xie, \& Bacon(1999)]{fan}
       Fan, J.H., Xie, G.Z., \& Bacon, R., 1999, \aaps, 136, 13
\bibitem[Ferrarese \& Merritt(2000)]{fm00}Ferrarese, L.,
        \& Merritt, D., 2000, \apj, 539, L9
\bibitem[Francis et al.(1991)]{f91}
       Francis, P.J., Hewett, P.C., Foltz, C.B., Chaffee, F.H.,
       Weymann, R.J., \& Morris, S.L., 1991, \apj, 373, 465
\bibitem[Gebhardt et al.(2000)]{g00}
       Gebhardt, K., et al., 2000, \apj, 539, L13
\bibitem[Georganopoulos, Kirk, \& Mastichiadis(2001)]{gkm01}
       Georganopoulos, M., Kirk, J.G., \& Mastichiadis, A.,
       2001, in ASP Conf. Ser. 227, Blazar Demographics and Physics, ed. P.
       Padovani \& C.M. Urry(San Francisco: ASP)
\bibitem[Ghisellini et al.(1998)]{gc98}
       Ghisellini, G., Celotti, A., Fossati, G., Maraschi, L.,
       \& Comastri, A., 1998, \mnras, 301, 451
\bibitem[Gu, Cao, \& Jiang (2001)]{g01}
       Gu, M., Cao, X., \& Jiang, D. R., 2001, \mnras, 327, 1111 
\bibitem[Gu \& Lu(2000)]{gl00}
       Gu, W.M., \& Lu, J.F., 2000, \apj, 540, L33 
\bibitem[Ho \& Kormendy(2000)]{hk00}Ho, L.C., \& Kormendy, J., 2000, The Encyclopedia of Astronomy and Astrophysics (Institute of Physics
    Publishing). (astro-ph/0003267)
\bibitem[Jackson \& Browne(1991)]{jb91}
    Jackson, N., \& Browne, I.W.A., 1991, \mnras, 250, 414
\bibitem[Kaspi et al.(2000)]{kas00}Kaspi, S., Smith, P.S., Netzer, H.,
    Maoz, D., Jannuzi, B.T., \& Giveon, U., 2000, \apj, 533, 631
\bibitem[Koratkar \& Blaes(1999)]{kb99}
    Koratkar, A., \& Blaes, O., 1999, \pasp, 111, 1
\bibitem[Laor(2000)]{laor00}
       Laor, A. 2000, \apj, 543, L111
\bibitem[Liu et al.(1999)]{l99}
      Liu, B.F., Yuan, W., Meyer, F.,  Meyer-Hofmeister, E., \& Xie, G.Z.,
      1999, \apj, 527, L17
\bibitem[Mahadevan(1997)]{m97}Mahadevan, R., 1997, \apj, 477, 585
\bibitem[Marcha et al. (1996)]{m96}
        Marcha, M.J.M., Browne, I.W.A., Impey, C.D., \& Smith, P.S.,
        1996, \mnras, 281, 425
\bibitem[McLure \& Dunlop(2001)]{md01}
     McLure, R.J., \& Dunlop, J.S., 2001, \mnras, 321, 515
\bibitem[Meyer \& Meyer-Hofmeister(1994)]{mm94}
      Meyer, F.,  \& Meyer-Hofmeister, E., 1994, \aap, 288, 175
\bibitem[Meyer-Hofmeister \& Meyer(2001)]{mm01}
     Meyer-Hofmeister, E., \& Meyer, F., 2001, \aap, 380, 744
\bibitem[Narayan(1996)]{n96}
       Narayan, R., 1996, \apj, 462, 136
\bibitem[Narayan \& Yi(1995)]{ny95}
       Narayan, R., \& Yi, I., 1995, \apj, 452, 710
\bibitem[Rawlings \& Saunders(1991)]{rs91}Rawlings, S.G., \& 
     Saunders, R.D.E., 1991, \nat, 349, 138
\bibitem[R\'oza\'nska \& Czerny(2000)]{rc00}
      R\'oza\'nska, A., \& Czerny, B., 2000,
        \aap, 360, 1170
\bibitem[Sitko \& Junkkarinen(1985)]{sj85}
       Sitko, M.L., \& Junkkarinen, V.T., 1985, \pasp, 97, 1158
\bibitem[Stickel, \& K\"uhr (1994)]{sk94}
      Stickel, M., \& K\"uhr, H., 1994, \aaps, 103, 349
\bibitem[Stickel, \& K\"uhr (1996)]{sk96}
      Stickel, M., \& K\"uhr, H., 1996, \aaps, 115, 1
\bibitem[Stickel, Meisenheimer, \& K\"uhr (1994)]{s94}
      Stickel, M., Meisenheimer, K., \& K\"uhr, H., 1994, \aaps,
      105, 211
\bibitem[Urry \& Padovani(1995)]{up95}
     Urry, C.M., \& Padovani, P., 1995, \pasp, 107, 803
\bibitem[Veron-Cetty \& Veron(2000)]{vv00}Veron-Cetty, M.P., \& Veron, P., 2000, \aapr, 10, 81
\bibitem[Wandel, Peterson, \& Malkan(1999)]{wpm99}Wandel, A.,
     Peterson, B.M., \& Malkan, M.A., 1999, \apj, 526, 579
\bibitem[Wang, Xue, \& Wang(2001)]{wang01}
       Wang, J.-M., Xue, S.J., \& Wang, J.C., 2001, \apj, submitted
\bibitem[Wu, Liu, \& Zhang(2002)]{wu02}
      Wu, X.B., Liu, F.K., Zhang, T.Z., 2002, \aap, accepted
      (astro-ph. 0203158)
\bibitem[Yi(1996)]{y96}Yi, I., 1996, \apj, 473, 645




\end{thebibliography}
\end{document}